\documentclass[prd,superscriptaddress,amsfonts,amssymb,amsmath,showpacs,twocolumn,floatfix]{revtex4-2}
\usepackage{bm}
\usepackage{amsfonts}
\usepackage{latexsym}
\usepackage{graphicx}
\usepackage{amsmath}
\usepackage{palatino}
\usepackage{mathpazo}
\usepackage{textcomp}
\usepackage{comment}
\usepackage{float}
\usepackage{amsmath,amssymb}

\usepackage{booktabs}
\usepackage{dcolumn}
\usepackage{booktabs}
\usepackage{multirow}
\usepackage{amsmath}
\usepackage{xcolor}
\usepackage{orcidlink}
\usepackage[caption=false]{subfig}
\usepackage{commath}
\usepackage{tabularx}

\def\jnl@style{\it}
\def\aaref@jnl#1{{\jnl@style#1}}

\def\aaref@jnl#1{{\jnl@style#1}}

\def\aj{\aaref@jnl{AJ}}                   
\def\apj{\aaref@jnl{ApJ}}                 
\def\apjl{\aaref@jnl{ApJ}}                
\def\apjs{\aaref@jnl{ApJS}}               
\def\apss{\aaref@jnl{Ap\&SS}}             
\def\aap{\aaref@jnl{A\&A}}                
\def\aapr{\aaref@jnl{A\&A~Rev.}}          
\def\aaps{\aaref@jnl{A\&AS}}              
\def\mnras{\aaref@jnl{Mon.~Not.~Roy.~Astron.~Soc.}}             
\def\prd{\aaref@jnl{Phys.~Rev.~D}}        
\def\prc{\aaref@jnl{Phys.~Rev.~C}}  
\def\prl{\aaref@jnl{Phys.~Rev.~Lett.}}    
\def\qjras{\aaref@jnl{QJRAS}}             
\def\skytel{\aaref@jnl{S\&T}}             
\def\ssr{\aaref@jnl{Space~Sci.~Rev.}}     
\def\zap{\aaref@jnl{ZAp}}                 
\def\nat{\aaref@jnl{Nature}}              
\def\aplett{\aaref@jnl{Astrophys.~Lett.}} 
\def\apspr{\aaref@jnl{Astrophys.~Space~Phys.~Res.}} 
\def\physrep{\aaref@jnl{Phys.~Rep.}}      
\def\physscr{\aaref@jnl{Phys.~Scr}}       
\def\commat{\aaref@jnl{Comm.~Math.~Phys.}}              
\def\science{\aaref@jnl{Science}}               
\def\cqg{\aaref@jnl{Classical Quant.~Grav.}}            
\def\jpcs{\aaref@jnl{JPCS}}                                     
\def\ijmpd{\aaref@jnl{Int.~J.~Mod.~Phys.~D}}                    
\def\grg{\aaref@jnl{Gen.~Relat.~Gravit.}}               
\def\rpp{\aaref@jnl{Rep.~Prog.~Phys.}}          
\def\npa{\aaref@jnl{Nucl.~Phys.~A}}        
\def\lrr{\aaref@jnl{Living Rev.~Rel.}}                   
\def\jcap{\aaref@jnl{J.~Cosmology Astropart.~Phys.}}    
\def\rmp{\aaref@jnl{Rev.~Mod.~Phys.}}   
\def\epjc{\aaref@jnl{Eur.~Phys.~J.~C}}

\allowdisplaybreaks[1]
\renewcommand{\arraystretch}{1.1}
\usepackage{hyperref}
\hypersetup{colorlinks,citecolor=blue}
\begin{document}

\color{black}       
\title{Observational tests of \texorpdfstring{$\Lambda(t)$}{Lambda(t)} cosmology in light of DESI DR2}

\author{D. Revanth Kumar\orcidlink{0009-0002-6599-3608}}
\email{2406c9m003@sru.edu.in}
\affiliation{Department of Mathematics, SR University, Warangal-506371, Telangana, India.}

\author{Santosh Kumar Yadav\orcidlink{0009-0009-2581-387X}}
\email{sky91bbaulko@gmail.com}
\affiliation{Department of Mathematics, SR University, Warangal-506371,
Telangana, India.}

\author{S. A. Kadam\orcidlink{0000-0002-2799-7870}}
\email{siddheshwar.kadam@dypiu.ac.in;
\\k.siddheshwar47@gmail.com}
\affiliation{Centre for Interdisciplinary Studies and Research, D Y Patil International University, Akurdi, Pune-411044, Maharashtra, India.}


\begin{abstract}
In this article, we investigate two phenomenological decaying vacuum cosmological models describing the accelerated expansion of the Universe. 
We constrain the model parameters using a Markov Chain Monte Carlo (MCMC) technique with recent datasets, including cosmic chronometer (CC), Pantheon+SH0ES (PPS), and DESI BAO data release (DR2).
Our analysis provides constraints from PPS, PPS+CC, and the joint PPS+CC+DR2 datasets for both models. All datasets favor $H_0 \simeq 72.53$--$73.01~\mathrm{Km\,s^{-1}\,Mpc^{-1}}$, while $\Omega_{m0}$ is higher with PPS alone and decreases to standard paradigm estimates with the inclusion of additional data. The evolution parameter is $n \approx 0.30$ from joint analysis, indicating a mild deviation from the $\Lambda$CDM framework.
Furthermore, the physical behavior of the models is examined through the deceleration parameter and the total equation of state, confirming a smooth transition from past deceleration expansion to the present accelerated expansion.

\textbf{Keywords:} Decaying vacuum cosmology, \texorpdfstring{$\Lambda(t)$}{Lambda(t)} cosmology, observational constraints, Hubble constant, dark energy.
\end{abstract}

\maketitle

\section{Introduction}\label{sec1}

Over the past two decades, a variety of independent astronomical observations have firmly established that the present Universe is undergoing an accelerated phase of expansion. This remarkable discovery was first revealed through observations of Type Ia supernovae~\cite{SupernovaSearchTeam:1998fmf,SupernovaCosmologyProject:1998vns} and later confirmed by several complementary probes, including cosmic microwave background (CMB) anisotropies~\cite{aghanim2020planck,WMAP:2003ivt}, baryon acoustic oscillations (BAO)~\cite{SDSS:2005xqv,SDSS:2009ocz}, and large-scale structure observations. Within the framework of general relativity, this late-time acceleration is commonly attributed to an exotic component with negative pressure, referred to as dark energy (DE), which dominates the present energy budget of the Universe~\cite{Peebles:2002gy,Padmanabhan:2002ji,Copeland:2006wr}. The simplest and well-accepted explanation for DE is the cosmological constant, denoted by $\Lambda$, leading to the standard $\Lambda$ cold dark matter ($\Lambda$CDM) model of the Universe. In this scenario, DE behaves as a constant vacuum energy density with an equation of state (EoS) parameter $\omega=-1$. 

The $\Lambda$CDM model successfully explains a wide range of cosmological observations, including the anisotropy spectrum of the CMB and the growth of large-scale structures~\cite{aghanim2020planck}. However, despite its observational success, the model faces several observational and theoretical challenges.
One of the major theoretical issues is the cosmological constant problem, which arises from the enormous discrepancy between the observed value of the vacuum energy density and the value predicted by quantum field theory~\cite{Weinberg:1988cp,Carroll:1991mt}. Another long-standing issue is the coincidence problem~\cite{Zlatev:1998tr}, which questions why the present-day energy densities of matter and DE are of comparable magnitude despite their different evolutionary behaviors throughout cosmic history.
Observationally, a significant discrepancy has been reported in the measurement of Hubble constant $H_0$. Late-time measurements favor larger values of $H_0$, while early-Universe estimates inferred from the CMB within the $\Lambda$CDM framework yield smaller values. This disagreement, known as the \textit{Hubble tension}, has been widely discussed in the literature~\cite{Planck:2018vyg,riess2022comprehensive,Verde:2019ivm,Knox:2019rjx,Kamionkowski:2022pkx,Li:2024yoe,Riess:2024vfa,Scolnic:2024hbh}, with the current significance exceeding $5\sigma$~\cite{riess2022comprehensive,Breuval:2024lsv}. Additional anomalies and observational inconsistencies have also been reported in recent studies~\cite{Rogers:2023upm,Ruchika:2024ymt,Lopez-Corredoira:2024pgl,Pourojaghi:2024bxa,Green:2024xbb,Khalife:2023qbu,Hazra:2024nav}. 
These theoretical and observational challenges have motivated numerous investigations into alternative cosmological scenarios that extend or modify the standard $\Lambda$CDM framework.

To explain cosmic acceleration, two broad classes of approaches have been widely explored in the literature. One approach involves modifying the gravitational sector of General Relativity, leading to a variety of modified gravity theories~\cite{Copeland:2006wr,Clifton:2011jh,Koyama:2015vza,nojiri2017modified,koussour2023constraining,duchaniya2024cosmological, Kadam:2026pjm}. The second approach introduces additional dynamical components in the cosmic energy budget, collectively referred to as DE~\cite{DeFelice:2010aj, Scherer:2025esj, Giare:2024gpk, Yadav:2019jio, nagpal2025late, Escamilla:2024fzq,Pacif:2020hai,
abdalla2022cosmology, RevanthKumar:2026ayb}. In this context, the equation of state parameter $\omega=p/\rho$ plays a central role in characterizing the physical properties of DE. Observational constraints suggest that the present value of $\omega$ is close to $-1$, consistent with accelerated expansion characterized by $\omega<-1/3$. Various theoretical models have been proposed to describe DE dynamics, including quintessence~\cite{Capozziello:2002rd, Zimdahl:2001ar, Amendola:1999er,Yadav:2024pvr,Koussour:2024jqp, Giare:2024gpk}, phantom models~\cite{Nojiri:2005sx, Caldwell:2003vq, Chiba:1999ka, Yang:2018prh, Koussour:2024kxd, Park:2025azv}, k-essence~\cite{Armendariz-Picon:2000nqq}, Chaplygin gas models~\cite{Kamenshchik:2001cp}, tachyon models~\cite{Sen:2002in}, and holographic DE scenarios~\cite{Li:2004rb}.

Motivated by the theoretical difficulties associated with a cosmological constant, several studies have explored the possibility that the vacuum energy density evolves with cosmic time, leading to dynamical vacuum or $\Lambda(t)$ cosmologies. In these scenarios, different phenomenological decay laws for $\Lambda(t)$ have been proposed, where the cosmological term depends on cosmological quantities such as the scale factor, cosmic time, or the Hubble parameter~\cite{47,48,69}. The evolution of the cosmological term has also been motivated by geometrical considerations and quantum mechanical arguments~\cite{50,51,52}. In addition, several works have examined possible interactions between vacuum energy and matter and confronted these models with cosmological observations~\cite{53,54,55}.  

Another important class of dynamical vacuum models is the running vacuum model, which emerges from the renormalization group approach of quantum field theory in curved spacetime. In this framework, the vacuum energy density can be expressed as a series expansion in powers of the Hubble parameter and its derivatives, where the leading correction beyond the constant term typically scales as $H^2$~\cite{57,58,59,60,61}. Further theoretical developments and reviews of this approach can be found in~\cite{64,65,66,67,68}. 
To address the cosmological constant problem, decaying vacuum models of the form $\Lambda \propto a^{-m}$ have been proposed, where $a$ is the scale factor and $m>0$. Quantum cosmological arguments often favor $m=2$, for which $\Lambda$ scales similarly to the curvature term~\cite{Lopez:1995eb,Chen:1990jw}. This behavior was first introduced by {\"O}zer and Taha as a possible resolution to several longstanding cosmological issues~\cite{Ozer:1986, Ozer:1987}.

Recent studies have continued to explore different realizations of $\Lambda(t)$ cosmologies and constrain them using observational data. In particular, dynamical vacuum models within the $\Lambda(t)$CDM framework have been investigated using observational probes, showing that such models can successfully describe the transition from a decelerated to an accelerated expansion phase of the Universe~\cite{Myrzakulov:2025pxy}. Power-law parametrizations of the cosmological term depending on the Hubble parameter, such as $\Lambda \propto H^n$, have also been examined using combined observational constraints~\cite{Myrzakulov:2024yfk}. Phenomenological models combining scale-factor and Hubble-dependent contributions to the vacuum term have likewise been analyzed and constrained within the $\Lambda(t)$CDM framework~\cite{Macedo:2023zrd}. More recently, interacting vacuum scenarios have been studied at both the background and perturbation levels using multiple cosmological observations~\cite{Escobal:2026lnp}. In addition, kinematic reconstructions of $\Lambda(t)$ models based on redshift parametrizations such as $\Lambda(z)=\Lambda_0+\Lambda_1 z$ have also been explored using observational datasets~\cite{Guillen:2025hix}.

Motivated by these works in literature, it is natural to consider phenomenological parametrizations of the cosmological constant that allow deviations from a constant value. 
In this work, we investigate a cosmological model characterized by a time-varying vacuum energy density $\Lambda(t)$ within the spatially flat Friedmann--Lemaître--Robertson--Walker (FLRW) Universe. We confront the proposed models with observational datasets, including cosmic chronometer measurements, Type Ia supernova data, and baryon acoustic oscillation observations from DESI DR2. We aim to constrain the model parameters and to examine whether a dynamical vacuum scenario provides a viable extension to the standard $\Lambda$CDM cosmology.

The rest of the paper is organized as follows: In Section~\ref{sec2}, we present the cosmological model equations and introduce the time-varying vacuum energy model considered in the present work. The observational datasets and statistical methodology followed to constrain the model parameters are described in Section~\ref{sec3}. The results obtained from the observational analysis are presented and discussed in Section~\ref{sec4}. Finally, Section~\ref{sec5} summarizes main findings and discusses their implications for the evolution of the Universe.

\section{Cosmological model}
\label{sec2}
We assume that the large-scale structure of the Universe satisfies the cosmological principle, according to which the Universe is homogeneous and isotropic on sufficiently large scales. 
Under this framework, the spacetime geometry is well-described by the Friedmann--Lemaître--Robertson--Walker (FLRW) metric,
\begin{equation}
ds^{2} = -c^{2}dt^{2} + a^{2}(t)\left[\frac{dr^{2}}{1 - kr^{2}} + r^{2}(d\theta^{2} + \sin^{2}\theta\, d\phi^{2})\right],
\end{equation}
where $a(t)$ is the scale factor describing the expansion of the Universe, and $k$ denotes the spatial curvature. In what follows, we adopt natural units such that $c = 1$.

The dynamics of the Universe are governed by Einstein’s field equations,
\begin{equation}
G_{\mu\nu} = 8\pi G T_{\mu\nu},
\end{equation}
where $T_{\mu\nu}$ is the energy-momentum tensor of the cosmic fluid. Considering a Universe filled with pressureless matter and vacuum energy, the total energy density and pressure can be written as
\begin{equation}
\rho_{\rm tot} = \rho_m + \rho_{\Lambda}, \quad p_{\rm tot} = p_m + p_{\Lambda}.
\end{equation}

In the late-time Universe, radiation can be neglected and matter is assumed to be pressureless ($p_m = 0$). The vacuum energy is characterized by an EoS parameter $\omega_\Lambda = -1$, which implies
\begin{equation}
p_\Lambda = -\rho_\Lambda.
\end{equation}

By considering these assumptions, Einstein’s equations reduce to the Friedmann equations,
\begin{align}
3H^{2} &= 8\pi G(\rho_m + \rho_{\Lambda}), \label{eq:f1}\\[6pt]
3\left(\dot{H} + H^2\right) &= -4\pi G\left(\rho_m - 2\rho_{\Lambda}\right), \label{eq:f2}
\end{align}
where $H = \dot{a}/a$ is the Hubble parameter.

In contrast to the standard $\Lambda$CDM model, we allow for a possible non-gravitational interaction between matter and vacuum energy. In this case, the conservation equation is modified as
\begin{equation}
\dot{\rho}_m + 3H\rho_m = -\dot{\Lambda}(t),
\end{equation}
which indicates that matter is not conserved independently and that the decaying vacuum can act as a source of matter.

Combining Eqs.~(\ref{eq:f1}) and (\ref{eq:f2}), we obtain the evolution equation governing the Hubble parameter. Multiplying Eq.~(\ref{eq:f2}) by 2 and adding it to Eq.~(\ref{eq:f1}), we obtain
\begin{equation}
3H^2 + 2\dot{H} = \Lambda.
\label{eq:main_evol}
\end{equation}

To facilitate comparison with observations, we express the evolution equations in terms of the redshift $z$ rather than cosmic time. The redshift is related to the scale factor through $1 + z = \frac{1}{a(t)}$.
Using this relation, the time derivative can be transformed as follows,
\begin{equation}
\frac{d}{dt} = -H(z)(1+z)\frac{d}{dz}.
\end{equation}

Applying this transformation to Eq.~(\ref{eq:main_evol}), we obtain the differential equation governing the evolution of the Hubble parameter,
\begin{equation}
\frac{dH}{dz} = \frac{3H}{2(1+z)} - \frac{\Lambda}{2H(1+z)}.
\label{eq:final}
\end{equation}

\section{Decaying Vacuum Model}\label{sec3}
We consider two phenomenological forms of a time-varying vacuum energy density.

\subsection{Case I: \texorpdfstring{$\Lambda(z) = \alpha (1+z)^n$}{Lambda(z) = alpha (1+z)\^n}}

We consider here a redshift-dependent parametrization of the vacuum energy density given by
\begin{equation}
\Lambda(z) = \alpha (1+z)^n,
\end{equation}
where $\alpha$ and $n$ are model parameters. At the present epoch ($z=0$), this relation reduces to $\Lambda_0 = \alpha$.

Substituting this form of $\Lambda(z)$ into Eq.~(\ref{eq:final}), the differential equation governing the Hubble parameter becomes
\begin{equation}
\frac{dH}{dz} = \frac{3H}{2(1+z)} - \frac{\alpha (1+z)^n}{2H(1+z)}.
\end{equation}

Solving the above equation, we obtain the following expression for the Hubble parameter:
\begin{equation}
H^2(z) = -\frac{\alpha}{n-3}(1+z)^n + H_0^2(1+z)^3 + \frac{\alpha}{n-3}(1+z)^3.
\label{model1}
\end{equation}

To express the solution in terms of observable cosmological parameters, we use the definition of the present-day vacuum density parameter,
\begin{equation}
\Omega_{\Lambda0} = \frac{\Lambda_0}{3H_0^2} = \frac{\alpha}{3H_0^2}.
\end{equation}
This gives $\alpha = 3H_0^2 \Omega_{\Lambda0}$. For a spatially flat Universe, we have $\Omega_{\Lambda0} = 1 - \Omega_{m0}$.

Substituting this relation into Eq.~(\ref{model1}), we can rewrite the Hubble parameter in a more convenient form as
\begin{equation}
H^2(z) = H_0^2 \left[ \beta(1+z)^n + (1 - \beta)(1+z)^3 \right],
\label{model_final}
\end{equation}
where the parameter $\beta$ is defined as
\begin{equation}
\beta = \frac{3(1-\Omega_{m0})}{3-n}.
\end{equation}

This expression clearly shows that the Hubble parameter consists of two contributions: one evolving as $(1+z)^n$, arising from the dynamical vacuum energy, and the other scaling as $(1+z)^3$, corresponding to the matter component.

\medskip

\subsection{Case II: \texorpdfstring{$\Lambda(z) = \alpha H^n$}{Lambda(z) = alpha H\^n}}

We now consider an alternative phenomenological model in which the vacuum energy density depends explicitly on the Hubble parameter in the form of a power law \cite{Myrzakulov:2024yfk},
\begin{equation}
\Lambda = \alpha H^n.
\end{equation}

Substituting this relation into Eq.~(\ref{eq:final}), we obtain the following differential equation
\begin{equation}
\frac{dH}{dz} = \frac{3H}{2(1+z)} - \frac{\alpha H^{\,n-1}}{2(1+z)}.
\end{equation}

Using the definition of the present-day density parameter, we have
\begin{equation}
\Omega_{\Lambda0} = \frac{\Lambda_0}{3H_0^2} = \frac{\alpha}{3H_0^{2-n}},
\end{equation}
which gives $\alpha = 3H_0^{2-n} \Omega_{\Lambda0}$.

Solving the above differential equation, the Hubble parameter can be expressed as
\begin{equation}
H(z) = H_0 \left[ \Omega_{m0} (1+z)^{3 - \frac{3n}{2}} + (1 - \Omega_{m0}) \right]^{\frac{1}{2-n}},
\label{model2}
\end{equation}

\medskip

Both models (Eqs.~\ref{model_final} and \ref{model2}) recover the standard $\Lambda$CDM limit for $n = 0$, where the vacuum energy density becomes constant.

\section{Data sets and methodology}
\label{sec4}

We employ recent cosmological observations, namely Cosmic Chronometer (CC) measurements of the Hubble function $H(z)$, Type Ia Supernovae (SNe Ia) distance moduli, and Baryon Acoustic Oscillation (BAO) measurements from the DESI Data Release 2 (DR2) to constrain the model parameters. The joint analysis of these largely independent datasets allows for robust constraints on cosmological parameters and enables consistency tests for cosmological scenarios.

In this analysis, estimation of parameters is performed using MCMC technique implemented with the \texttt{emcee} Python package. We consider identical flat (uniform) priors on all free parameters for both models, given by $40 < H_{0} < 100$, $0.01 < \Omega_{m0} < 0.9$, and $-4 < n < 4$. The convergence of the MCMC sample chains is assessed using the Gelman-Rubin criterion ($R-1 < 0.01$), and the obtained posterior distributions are analyzed using the publicly available \texttt{GetDist} package~\cite{Lewis:2019xzd}. A brief description of the datasets utilized here is mentioned below.

\subsection{Cosmic Chronometer (CC)}

The Cosmic Chronometer (CC) technique provides a direct, model-independent measurement of the Hubble parameter $H(z)$ based on the differential ages of passively evolving galaxies~\cite{Simon:2004tf,Stern:2009ep,Moresco:2015cya,Ratsimbazafy:2017vga}.  
We analyze 31 CC data points across the redshift interval $0 < z < 1.96$, as summarized in Table~\ref{CC}.
\begin{table}[ht]
    \centering
    \caption{The 31 $H(z)$ measurements from the CC method used in this study in units of $\mathrm{Km\,s^{-1} Mpc^{-1}}$.}
    \label{CC}
        \centering
        \setlength{\tabcolsep}{9pt}
        \begin{tabular}{c c c c}
\hline\hline
$z$ & $H(z)$ & $\sigma_H$ & Reference \\
\hline

0.07  & 69.0  & 19.6 & \multirow{4}{*}{\cite{zhang2016test}} \\
0.12  & 68.6  & 26.2 & \\
0.20  & 72.9  & 29.6 & \\
0.28  & 88.8  & 36.6 & \\
\hline
0.09  & 69.0  & 12.0 & \multirow{9}{*}{\cite{Simon:2004tf}} \\
0.17  & 83.0  & 8.0  & \\
0.27  & 77.0  & 14.0 & \\
0.40  & 95.0  & 17.0 & \\
0.9   & 117.0 & 23.0 & \\
1.3   & 168.0 & 17.0 & \\
1.43  & 177.0 & 18.0 & \\
1.53  & 140.0 & 14.0 & \\
1.75  & 202.0 & 40.0 & \\
\hline
0.179 & 75.0  & 4.0  & \multirow{8}{*}{\cite{Moresco:2012jh}} \\
0.199 & 75.0  & 5.0  & \\
0.352 & 83.0  & 14.0 & \\
0.593 & 104.0 & 13.0 & \\
0.68  & 92.0  & 8.0  & \\
0.781 & 105.0 & 12.0 & \\
0.875 & 125.0 & 17.0 & \\
1.037 & 154.0 & 20.0 & \\
\hline
0.3802 & 83.0 & 13.5 & \multirow{5}{*}{\cite{moresco20166}} \\
0.4004 & 77.0 & 10.2 & \\
0.4247 & 87.1 & 11.2 & \\
0.4497 & 92.8 & 12.9 & \\
0.4783 & 80.9 & 9.0  & \\
\hline
0.48  & 97.0 & 62.0 & \multirow{2}{*}{\cite{Stern:2009ep}} \\
0.88  & 90.0 & 40.0 & \\
\hline
0.47  & 89.0 & 50.0 & \cite{Ratsimbazafy:2017vga} \\
\hline
1.363 & 160.0 & 33.6 & \multirow{2}{*}{\cite{Moresco:2015cya}} \\
1.965 & 186.5 & 50.4 & \\

\hline\hline
\end{tabular}
\end{table}

Each data point provides an observed value $H_{\text{obs}}(z_{i})$ and an associated uncertainty $\sigma_{H,i}$. The chi-squared estimator for the CC dataset is defined as
\begin{equation}
\chi^{2}_{\text{CC}} = \sum_{i=1}^{N_{\text{CC}}} \frac{\left[H_{\text{th}}(z_{i}; \boldsymbol{\theta}) - H_{\text{obs}}(z_{i})\right]^{2}}{\sigma_{H,i}^{2}},
\end{equation}
where $H_{\text{th}}(z_{i}; \boldsymbol{\theta})$ is the theoretical prediction for the Hubble parameter from the model under consideration, and $\boldsymbol{\theta}$ denotes the parameter vector depending on the model.

\subsection{Type Ia Supernovae}
Type Ia supernovae (SNe Ia) are widely used as standard candles due to their nearly uniform peak luminosity, making them powerful probes of the late-time cosmic expansion. In this work, we utilize the distance modulus measurements from the Pantheon+SH0ES (\texttt{PPS}) compilation~\cite{brout2022pantheon+}. This dataset comprises 1701 light curves corresponding to 1550 spectroscopically confirmed SNe Ia events, spanning the redshift range $[0.001,~ 2.26]$.

The theoretical apparent magnitude of a supernova at redshift $z$ is given by
\begin{equation}
m_B(z) = 5 \log_{10} \left[ \frac{d_L(z)}{\mathrm{Mpc}} \right] + 25 + M_B,
\end{equation}
where the absolute magnitude is denoted by $M_B$ and the luminosity distance by $d_L(z)$. The latter is defined as
\begin{equation}
d_L(z) = {(1+z)} \int_{0}^{z} \frac{dx}{H(x)},
\end{equation}
The corresponding theoretical distance modulus is given by
\begin{equation}
\mu_{\mathrm{th}}(z) = 5 \log_{10} \left[ \frac{d_L(z)}{\mathrm{Mpc}} \right] + 25,
\end{equation}
which is directly compared with the observed distance modulus $\mu_{\mathrm{obs}}$.

The likelihood for the \texttt{PPS} dataset is constructed using the chi-squared statistic
\begin{equation}
\chi^2_{\mathrm{PPS}} = \Delta\boldsymbol{\mu}^{T} \mathbf{C}_{\mathrm{SN}}^{-1} \Delta\boldsymbol{\mu},
\end{equation}
where $\Delta\boldsymbol{\mu} = \mu_{\mathrm{th}} - \mu_{\mathrm{obs}}$ is the residual vector, and $\mathbf{C}_{\mathrm{SN}}$ is the full covariance matrix incorporating both statistical and systematic uncertainties.

\subsection{DESI BAO DR2 Dataset}


Baryon acoustic oscillations (BAO) are periodic features in the baryonic matter distribution that act as standard rulers in cosmology. The BAO scale is characterized by the comoving sound horizon at the drag epoch, $r_d$, defined as the distance travelled by sound waves in the baryon-photon fluid before decoupling effects cease.
The sound horizon is given by
\begin{equation}
  r_d = \int_{z_d}^{\infty} \frac{c_s(z)}{H(z)} \, dz ,
\end{equation}
where $c_s(z)$ is the sound speed and $z_d$ is the drag epoch redshift.

The DESI is a spectroscopic survey that enhances BAO constraints by probing the large-scale structure. It employs multiple matter tracers to measure BAO over a range from $0.3$ to $2.33$. The isotropic BAO datasets provide measurements 
from Bright Galaxy Survey (BGS). The anisotropic BAO measurements come from Luminous Red Galaxies (LRGs), Emission Line Galaxies (ELGs), and Quasi-Stellar Objects (QSOs), and Lyman-$\alpha$ quasars (Lya QSOs). 
In the second data release (DR2), the DESI survey has been significantly increased, covering 6671 dark tiles and 5171 bright tiles. This corresponds to an improvement by a factor of about 2.4 for the dark program and 2.3 for the bright program compared to DR1~\cite{DESI:2024mwx}. Using the BAO measurements from DESI DR2, we can constrain the combined parameter $H_0 r_d$, but not $H_0$ or $r_d$ individually.
In this work, we have considered DESI BAO DR2 observations from~\cite{karim2025desi} as mentioned here in Table~\ref{DESI BAO}. We refer to this data simply as \texttt{DR2} in the remaining text.

\begin{table}[ht]
\caption{ The 9 points from DESI BAO DR2 measurements used in the present analysis in units of $\mathrm{Km\,s^{-1} Mpc^{-1}}$.}\label{DESI BAO}

\setlength{\tabcolsep}{4pt}

\begin{tabular}{l c c c c}
\hline\hline
$z_{\rm eff}$ & $D_M/r_d$ & $D_H/r_d$ & $D_V/r_d$ \\
\hline
$0.295$ & $-$ & $-$ & $7.942 \pm 0.075$ \\
$0.510$ & $13.588 \pm 0.167$ & $21.863 \pm 0.425$ & $12.720 \pm 0.099$ \\
$0.706$ & $17.351 \pm 0.177$ & $19.455 \pm 0.330$ & $16.050 \pm 0.110$ \\
$0.934$ & $21.576 \pm 0.152$ & $17.641 \pm 0.193$ & $19.721 \pm 0.091$ \\
$0.922$ & $21.648 \pm 0.178$ & $17.577 \pm 0.213$ & $19.656 \pm 0.105$ \\
$0.955$ & $21.707 \pm 0.335$ & $17.803 \pm 0.297$ & $20.008 \pm 0.183$ \\
$1.321$ & $27.601 \pm 0.318$ & $14.176 \pm 0.221$ & $24.252 \pm 0.174$ \\
$1.484$ & $30.512 \pm 0.760$ & $12.817 \pm 0.516$ & $26.055 \pm 0.398$ \\
$2.330$ & $38.988 \pm 0.531$ & $8.632 \pm 0.101$ & $31.267 \pm 0.256$ \\
\hline\hline
\end{tabular}
\end{table}
The chi-squared statistic for the BAO dataset is given by
\begin{align}
\chi^2_{\rm DR2} = \sum_{i} \left[ 
\left( \frac{\Delta D_{M,i}}{\sigma_{D_M,i}} \right)^2 +
\left( \frac{\Delta D_{H,i}}{\sigma_{D_H,i}} \right)^2 +
\left( \frac{\Delta D_{V,i}}{\sigma_{D_V,i}} \right)^2
\right],
\end{align}
where $\Delta \mathbf{D}$ represents the difference between observed and theoretical values of the BAO observables, and $\sigma$ represents the corresponding uncertainties.


Assuming that the three datasets are statistically independent, the combined likelihood can be constructed as
\begin{equation*}
\mathcal{L}_{\text{tot}} \propto \exp\left[-\frac{1}{2}\left(\chi^{2}_{\text{CC}} + \chi^{2}_{\text{SNe Ia}} + \chi^{2}_{\text{DR2}}\right)\right].
\end{equation*}
Equivalently, the total chi-squared for the combined likelihood is given by
\begin{equation*}
\chi^2_{\rm tot} = \chi^2_{\rm CC} + \chi^2_{\rm SN} + \chi^2_{\rm DR2}.
\end{equation*}

\section{Results and Discussion}\label{sec5}

Here, we have interpreted the observational constraints on the model parameters obtained from the PPS, PPS+CC, and combined analysis PPS+CC+DR2 within the framework of the proposed decaying vacuum models. In this scenario, deviations from the standard $\Lambda$CDM cosmology are governed by the parameter $n$, which controls the redshift evolution of the vacuum energy density through the relation $\Lambda(z) \propto (1+z)^n$ and $\Lambda \propto H^n$. The limit $n=0$ corresponds to a constant vacuum energy, recovering the $\Lambda$CDM case, while non-zero values of $n$ introduce dynamical evolution in the DE sector.
The marginalized constraints on the parameters $H_0$, $\Omega_{m0}$, and $n$ for both Model 1 and Model 2 are summarized in Table~\ref{tabparams}. 

\begin{table*}[htbp]
\centering
\caption{Observational constraints on model parameters ($H_0$, $\Omega_{m0}$, $n$) for Model 1 and Model 2 from different dataset combinations at 68\%, 95\% and 99\% confidence levels.}
\label{tabparams}
\renewcommand{\arraystretch}{2}
\setlength{\tabcolsep}{6pt}

\begin{tabular}{l c c c c}
\hline\hline
Dataset & Model & $H_0$ {\footnotesize($Km \ s^{-1} \ Mpc^{-1}$)} & $\Omega_{m0}$ & $n$ \\
\hline

\multirow{2}{*}{PPS}
& Model 1 
& $72.53 \pm 0.28^{+0.55+0.72}_{-0.53-0.69}$ 
& $0.435^{+0.045+0.079+0.097}_{-0.037-0.083-0.12}$ 
& $0.78^{+0.44+0.75+0.88}_{-0.34-0.79-1.2}$ \\

& Model 2 
& $72.53 \pm 0.28^{+0.54+0.73}_{-0.54-0.71}$ 
& $0.434 \pm 0.042^{+0.082+0.11}_{-0.084-0.11}$ 
& $1.06^{+0.57+1.0+1.3}_{-0.49-1.1-1.5}$ \\
\hline

\multirow{2}{*}{PPS + CC}
& Model 1 
& $72.64 \pm 0.27^{+0.52+0.69}_{-0.51-0.67}$ 
& $0.425 \pm 0.034^{+0.064+0.082}_{-0.069-0.092}$ 
& $0.80^{+0.28+0.49+0.63}_{-0.24-0.55-0.74}$ \\

& Model 2 
& $72.65 \pm 0.27^{+0.53+0.69}_{-0.52-0.68}$ 
& $0.419 \pm 0.035^{+0.068+0.089}_{-0.068-0.090}$ 
& $1.04^{+0.37+0.67+0.88}_{-0.33-0.71-0.95}$ \\
\hline

\multirow{2}{*}{PPS + CC + DR2}
& Model 1 
& $72.93 \pm 0.23^{+0.45+0.60}_{-0.44-0.58}$ 
& $0.368 \pm 0.024^{+0.047+0.061}_{-0.048-0.063}$ 
& $0.30 \pm 0.12^{+0.23+0.30}_{-0.23-0.31}$ \\

& Model 2 
& $73.01 \pm 0.26^{+0.43+0.57}_{-0.44-0.58}$ 
& $0.354 \pm 0.022^{+0.043+0.057}_{-0.042-0.055}$ 
& $0.30 \pm 0.14^{+0.24+0.32}_{-0.26-0.34}$ \\
\hline\hline

\end{tabular}
\end{table*}

For the PPS dataset, both models yield consistent estimates of the Hubble constant, $H_0 = 72.53 \pm 0.28\,\mathrm{Km\,s^{-1}\,Mpc^{-1}}$ at $68\%$ confidence level, which is in close agreement with the local SH0ES measurement, $H_0 = 73.04 \pm 1.04\,\mathrm{Km\,s^{-1}\,Mpc^{-1}}$~\cite{SupernovaSearchTeam:2004lze}.
The matter density parameter is constrained to $\Omega_{m0} \approx 0.435$ for Model 1 and $\Omega_{m0} \approx 0.434$ for Model 2. The corresponding constraints on the evolution parameter show moderate values, with $n = 0.78^{+0.44}_{-0.34}$ for Model 1 and a comparatively higher value $n = 1.06^{+0.57}_{-0.49}$ for Model 2. The larger value of $n$ in Model 2 indicates a relatively stronger redshift evolution of the vacuum energy component.

The inclusion of cosmic chronometer (CC) data in the PPS+CC dataset leads to improved constraints without significantly altering the mean values of the parameters. The Hubble constant slightly increases to $H_0 = 72.64 \pm 0.27\,\mathrm{Km\,s^{-1}\,Mpc^{-1}}$ for Model 1 and $72.65 \pm 0.27\,\mathrm{Km\,s^{-1}\,Mpc^{-1}}$ for Model 2. At the same time, the matter density decreases marginally to $\Omega_{m0} = 0.425$ (Model 1) and $0.419$ (Model 2), while the parameter $n$ remains nearly unchanged, with $n = 0.80^{+0.28}_{-0.24}$ and $n = 1.04^{+0.37}_{-0.33}$, respectively. This behavior suggests that CC data provide complementary constraints on the expansion history, reducing uncertainties while preserving the overall parameter trends.

The most stringent constraints are obtained from the combined PPS+CC+DR2 dataset. The inclusion of DR2 data significantly tightens the parameter space and leads to noticeable shifts in the inferred values. The Hubble constant increases slightly to $H_0 = 72.93 \pm 0.23\,\mathrm{Km\,s^{-1}\,Mpc^{-1}}$ for Model 1 and $73.01 \pm 0.26\,\mathrm{Km\,s^{-1}\,Mpc^{-1}}$ for Model 2, remaining consistent with local measurements. Meanwhile, the matter density parameter decreases more substantially to $\Omega_{m0} = 0.368$ and $0.354$, respectively. Most importantly, the evolution parameter is strongly constrained to lower values, $n \approx 0.30$ for both models, with significantly reduced uncertainties. This indicates that large deviations from a constant vacuum energy are disfavored when multiple observational probes are combined.

From a physical perspective, the reduction in $n$ in the joint analysis implies that the vacuum energy evolves only mildly with redshift, remaining close to the $\Lambda$CDM scenario at late times. Although $n$ remains positive across all dataset combinations, indicating a departure from a strictly constant vacuum energy.

\begin{figure}[ht]
    \centering
        \includegraphics[width=1.0\linewidth]{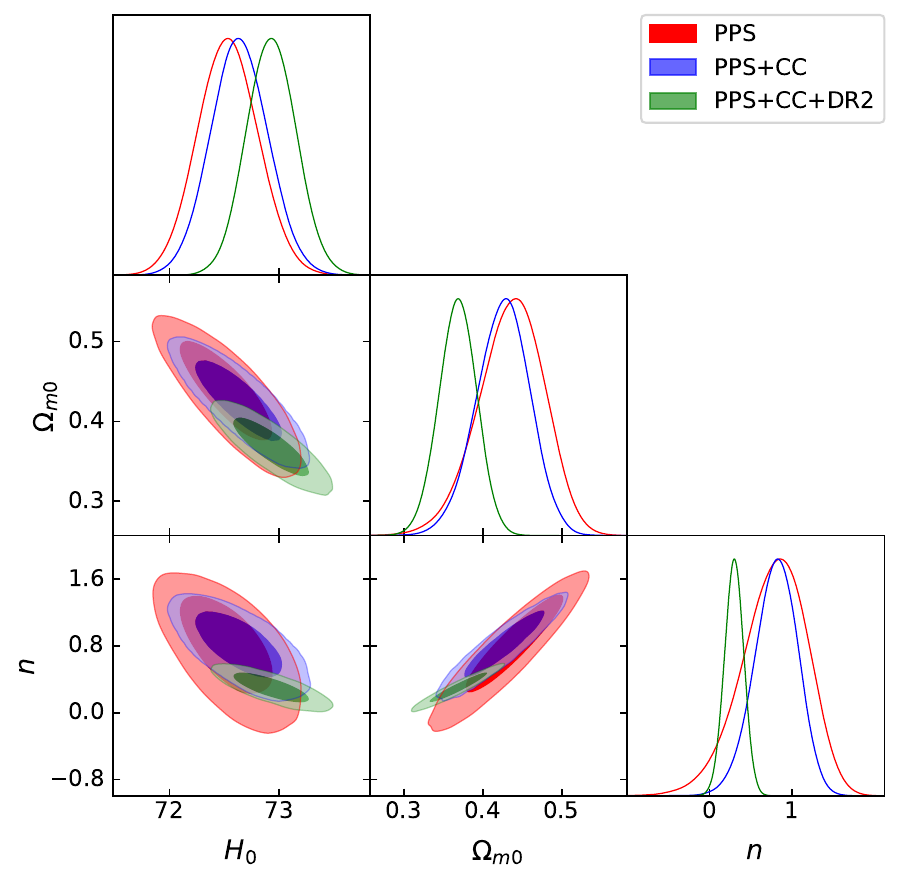}
        \caption{Contour plots indicating $1\sigma$ and $2\sigma$ confidence regions for model 1 parameters for considered datasets.}
        \label{fig:1a}
\end{figure}

\begin{figure}[ht]
        \centering
        \includegraphics[width=1.0\linewidth]{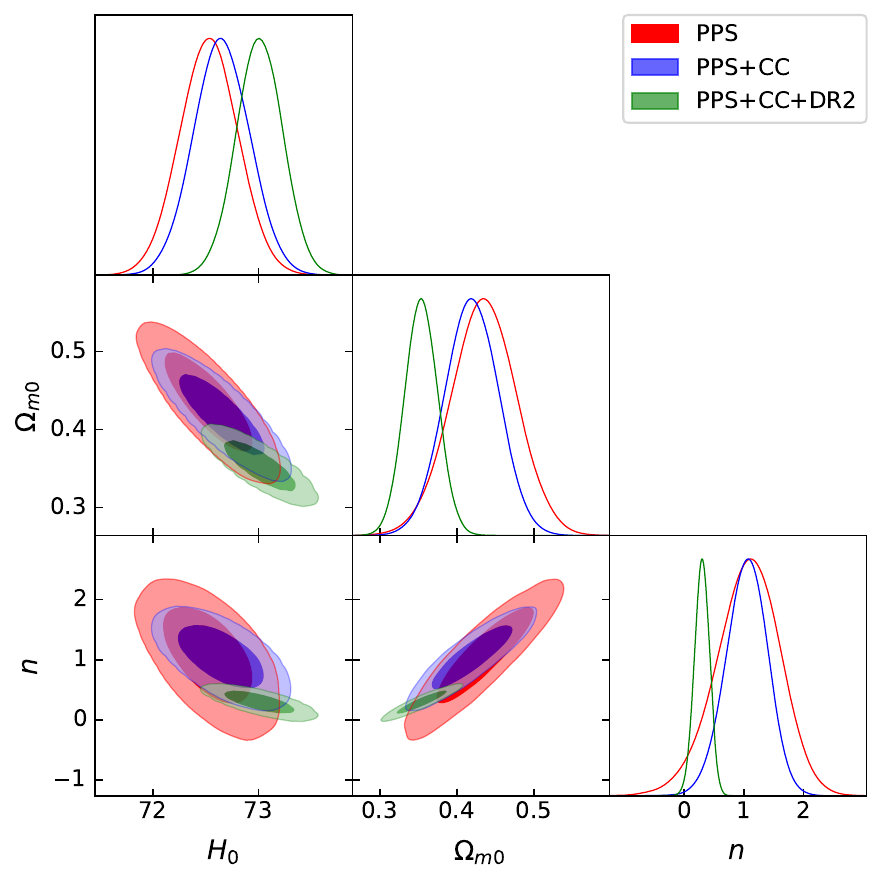}
        \caption{Contour plots indicating $1\sigma$ and $2\sigma$ confidence regions for model 2 parameters for considered datasets.}
        \label{fig:1b}
\end{figure}

The two-dimensional contour plots shown in Fig.~\ref{fig:1a} and Fig.~\ref{fig:1b} illustrate the correlations among the model parameters. A clear positive correlation between $\Omega_{m0}$ and $n$ is observed for all datasets and both models, indicating that higher matter density favors stronger evolution of the vacuum energy component. This degeneracy reflects the interplay between matter and vacuum energy in shaping the expansion history of the Universe.

We also observe a systematic decrease in the matter density parameter as additional datasets are included. For Model 1, $\Omega_{m0}$ decreases from $0.435$ (PPS) to $0.368$ (PPS+CC+DR2), while for Model 2 it decreases from $0.434$ to $0.354$. This trend reflects the role of combined datasets in breaking degeneracies and tightening parameter constraints. The shrinking of the one-dimensional marginalized distributions and the progressively tighter contour regions further confirm the enhanced constraining power of the joint analysis.

Overall, the results obtained in this work are consistent across different dataset combinations and between the two models. The combined PPS+CC+DR2 dataset provides the strongest constraints, favoring a cosmological scenario with lower matter density and a mildly evolving vacuum energy characterized by a small positive value of $n$. These findings indicate that the proposed decaying vacuum models remain consistent with current observational data and can serve as viable extensions to the standard $\Lambda$CDM framework.

\subsection{Statistical Analysis}
To assess the goodness of fit of the proposed models with the observational datasets, we perform a chi-square ($\chi^2$) analysis. The minimum chi-square, $\chi^2_{\text{min}}$, is obtained from the maximum likelihood estimation through the relation $\chi^2_{\text{min}} = -2\ln(\mathcal{L}{\text{max}})$, where $\mathcal{L}{\text{max}}$ denotes the maximum likelihood function. The reduced chi-square, $\chi^2_r$, is then defined by normalizing $\chi^2_{\text{min}}$ with the number of degrees of freedom, i.e., $\chi^2_r = \chi^2_{\text{min}}/(N - k)$, where $N$ represents the total number of data points and $k$ is the number of free parameters in the model.
In general, a value of $\chi^2_r \approx 1$ indicates a good agreement between the model and observational data, while significant deviations from unity may imply underfitting ($\chi^2_r \gg 1$) or overfitting ($\chi^2_r \ll 1$). The estimated values of $\chi^2_{\text{min}}$ and $\chi^2_r$ for different dataset combinations are presented in Table~\ref{tab:physical}. 

\begin{table*}[ht]
\centering
\caption{Observational constraints on physical parameters ($q_0$, $z_{tr}$, $\omega_0$ along with $\chi^2_{min}$ and $\chi^2_{r}$ for Model 1 and Model 2 from different dataset combinations.}
\label{tab:physical}
\renewcommand{\arraystretch}{1.5}
\setlength{\tabcolsep}{6pt}
\begin{tabular}{l c c c c c c}
\hline\hline
Dataset & Model & $q_0$ & $z_{tr}$ & $\omega_0$ & $\chi^2_{min}$ & $\chi^2_{r}$\\
\hline
\multirow{2}{*}{PPS}
& Model 1 
& $-0.3475$ & $0.8543$ & $-0.565$ &1749 &1.03 \\
& Model 2 
& $-0.3490$ & $0.9737$ & $-0.566$ &1749 &1.03 \\
\hline
\multirow{2}{*}{PPS + CC}
& Model 1 
& $-0.3625$ & $0.9524$ & $-0.575$ &1774 &1.03 \\
& Model 2 
& $-0.3715$ & $1.0306$ & $-0.581$ &1774 &1.03 \\
\hline
\multirow{2}{*}{PPS + CC + DR2}
& Model 1 
& $-0.4480$ & $0.6724$ & $-0.632$ &1796 &1.03 \\
& Model 2 
& $-0.4690$ & $0.6615$ & $-0.646$ &1797 &1.03 \\
\hline\hline
\end{tabular}
\end{table*}
The values $\chi^2_r \approx 1.03$ across all dataset combinations indicate a statistically consistent and satisfactory fit, with no evidence of underfitting or overfitting. The small variation in $\chi^2_{\text{min}}$ between the two models further suggests that they are equally compatible with the observational data. Therefore, from a statistical perspective, neither model is strongly favored with the datasets used, and both remain viable alternatives to the standard $\Lambda$CDM cosmology.

\subsection{Behaviour of Deceleration Parameter}
The expansion dynamics of the Universe can be effectively described through the deceleration parameter, which quantifies the rate at which the cosmic expansion accelerates or decelerates. It is defined as
\begin{equation}
q = -1 - \frac{\dot{H}}{H^2},
\end{equation}
where the overdot denotes derivative with respect to cosmic time. A positive value of $q$ indicates a decelerating phase dominated by matter or radiation, while a negative value signifies an accelerated expansion driven by DE.

Using the expressions of the Hubble parameter derived for Model 1~\eqref{model_final} and Model 2~\eqref{model2}, the corresponding deceleration parameters $q(z)$ are obtained as:
\paragraph{Model 1:}
\begin{equation}
q(z) = \frac{A(n-2) + B}{2(A+B)},
\end{equation}
where
\begin{align*}
A = \frac{3(1-\Omega_{m0})(1+z)^n}{3-n}, \\
B = \left(1 - \frac{3(1-\Omega_{m0})}{3-n}\right)(1+z)^3.
\end{align*}

\begin{figure}[ht]
        \centering
        \includegraphics[width=1\linewidth]{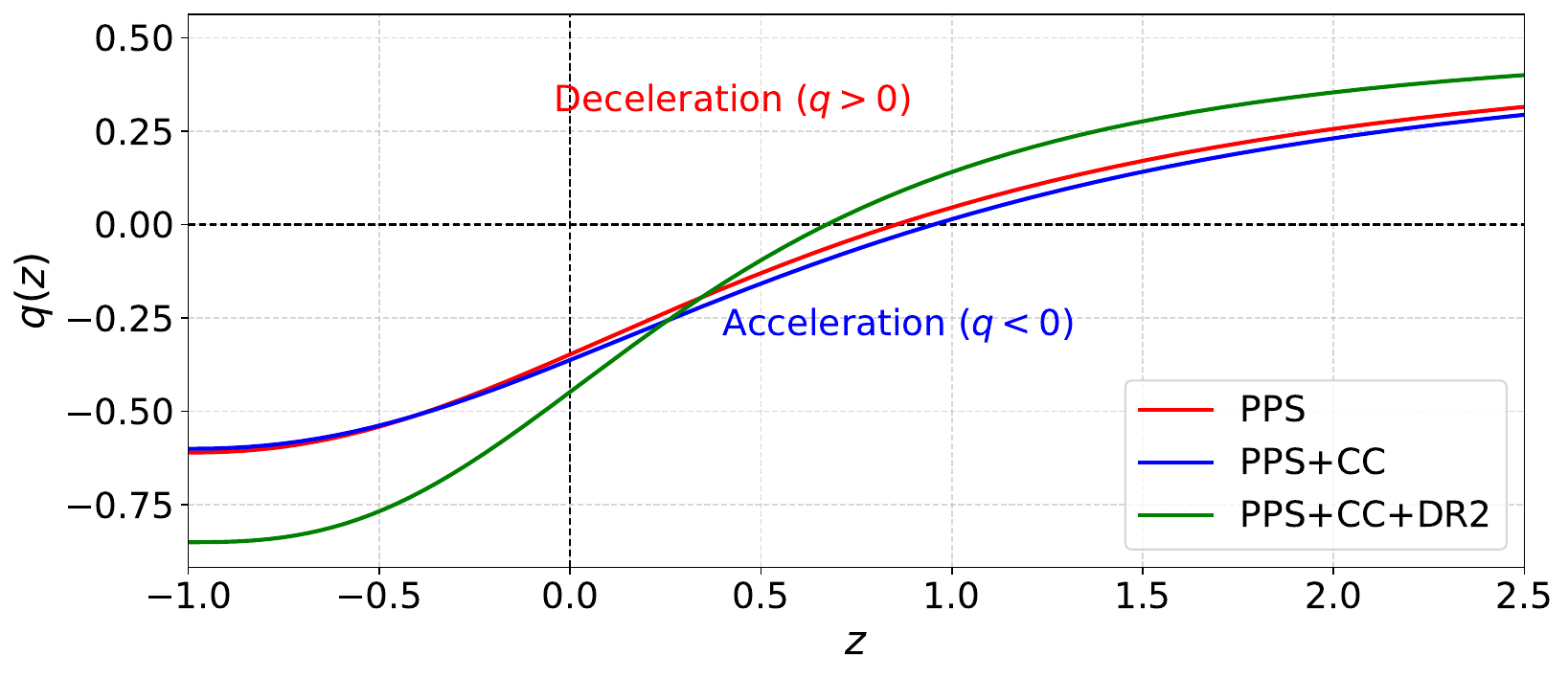}
        \caption{Deceleration parameter $q(z)$ vs redshift $z$ for Model 1.}
        \label{fig:2a}
    \end{figure}

\begin{figure}[ht]
        \centering
        \includegraphics[width=1\linewidth]{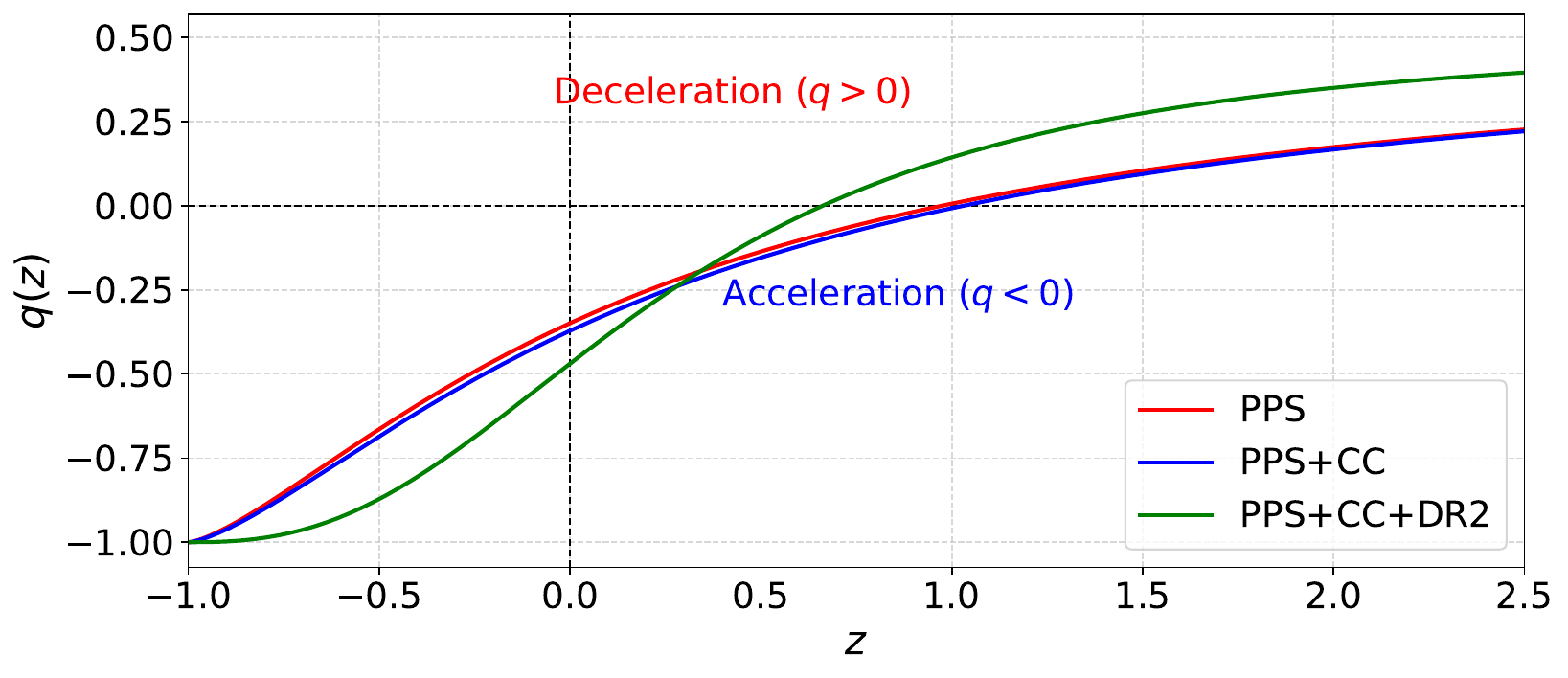}
        \caption{Deceleration parameter $q(z)$ vs redshift $z$ for Model 2.}
        \label{fig:2b}
    \end{figure}

\paragraph{Model 2:}
\begin{equation}
q(z) = -1 - \frac{\left(3 - \frac{3n}{2}\right)\Omega_{m0}(1+z)^{\left(3 - \frac{3n}{2}\right)}}{(n-2)\left[\Omega_{m0}(1+z)^{\left(3 - \frac{3n}{2}\right)} - 1\right] + 1}.
\end{equation}
The evolution of $q(z)$ as a function of redshift is presented in Fig.~\ref{fig:2a} and Fig.~\ref{fig:2b} for Model 1 and Model 2, respectively, over the range $z \in [-1, 2.5]$. The plots clearly show a transition from a decelerating phase ($q>0$) at earlier epochs to an accelerating phase ($q<0$) at late times.

The present-day values of the deceleration parameter ($q_0$) and the transition redshift ($z_{\mathrm{tr}}$), at which the Universe switches from deceleration to acceleration, are summarized in Table~\ref{tab:physical}. For Model 1, the present value evolves from $q_0 = -0.3475$ (PPS) to $q_0 = -0.4480$ (PPS+CC+DR2), while for Model 2 it varies from $q_0 = -0.3490$ to $q_0 = -0.4690$ across the same datasets. These negative values of $q_0$ confirm that the Universe is currently in an accelerated expansion phase. The obtained results from the joint analysis are consistent with the recent literature as $q_0=-0.41$ by Iqbal et al.~\cite{iqbal2025exploring}, $q_0=-0.44$ by Bhagat et al.~\cite{bhagat2025exploring}, $q_0=-0.43$ by Wang et al.~\cite{Wang:2023iuj}.

The transition redshift is found to lie in the range $z_{\mathrm{tr}} \approx 0.85$--$1.03$ for PPS and PPS+CC datasets, and decreases to $z_{\mathrm{tr}} \approx 0.66$--$0.67$ when DR2 data are included, which are consistent with $z_{\mathrm{tr}} \approx 0.65$ by Myrzakulov et.~\cite{Myrzakulov:2025pxy} and Koussour et al.~\cite{Koussour:2023nhw}. 
The smooth evolution of $q(z)$ across all datasets indicates that the models provide a stable and reliable description of the late-time expansion history.

Furthermore, the behavior of $q(z)$ in the future region ($z<0$) suggests that the accelerated expansion persists without any indication of a transition back to deceleration within the considered redshift range. Overall, these results are consistent with the standard cosmological picture of a transition from past deceleration to present acceleration.

\subsection{Total EoS and Energy Density}

In this study, we assume that DE corresponds to vacuum energy with equation of state $\omega_{\Lambda} = -1$. We now investigate the total equation of state parameter, which accounts for the combined contribution of matter and vacuum energy. It is given by
\begin{equation}
\omega_{\mathrm{tot}} = \frac{p_{\Lambda}}{\rho_m + \rho_{\Lambda}} = -\frac{1}{1 + \frac{\rho_m}{\rho_{\Lambda}}},
\end{equation}
which clearly shows that the evolution of $\omega_{\mathrm{tot}}$ is governed by the ratio of matter to vacuum energy densities.

The EoS parameter plays a key role in distinguishing different phases of cosmic evolution. In particular, $\omega = 0$ corresponds to a matter-dominated phase, $\omega = \frac{1}{3}$ to radiation domination, and $\omega = -1$ represents a vacuum energy dominated Universe. The condition $\omega_{\mathrm{tot}} < -\frac{1}{3}$ indicates accelerated expansion, while $-1 < \omega_{\mathrm{tot}} < -\frac{1}{3}$ corresponds to the quintessence regime.

\begin{figure}[ht]
        \centering
        \includegraphics[width=\linewidth]{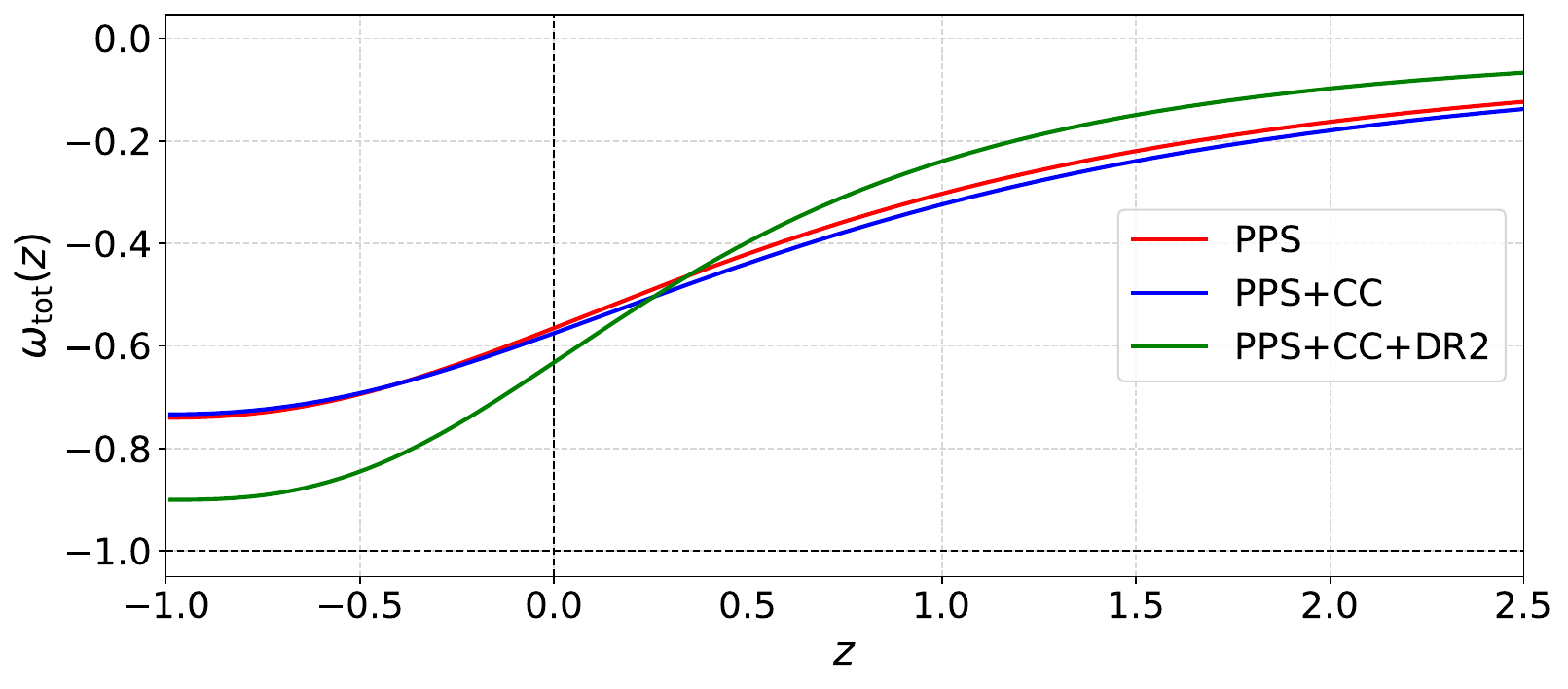}
        \caption{Total equation of state $\omega_{tot}$ vs redshift $z$ for Model 1.}
        \label{fig:3a}
    \end{figure}

\begin{figure}[ht]
        \centering
        \includegraphics[width=\linewidth]{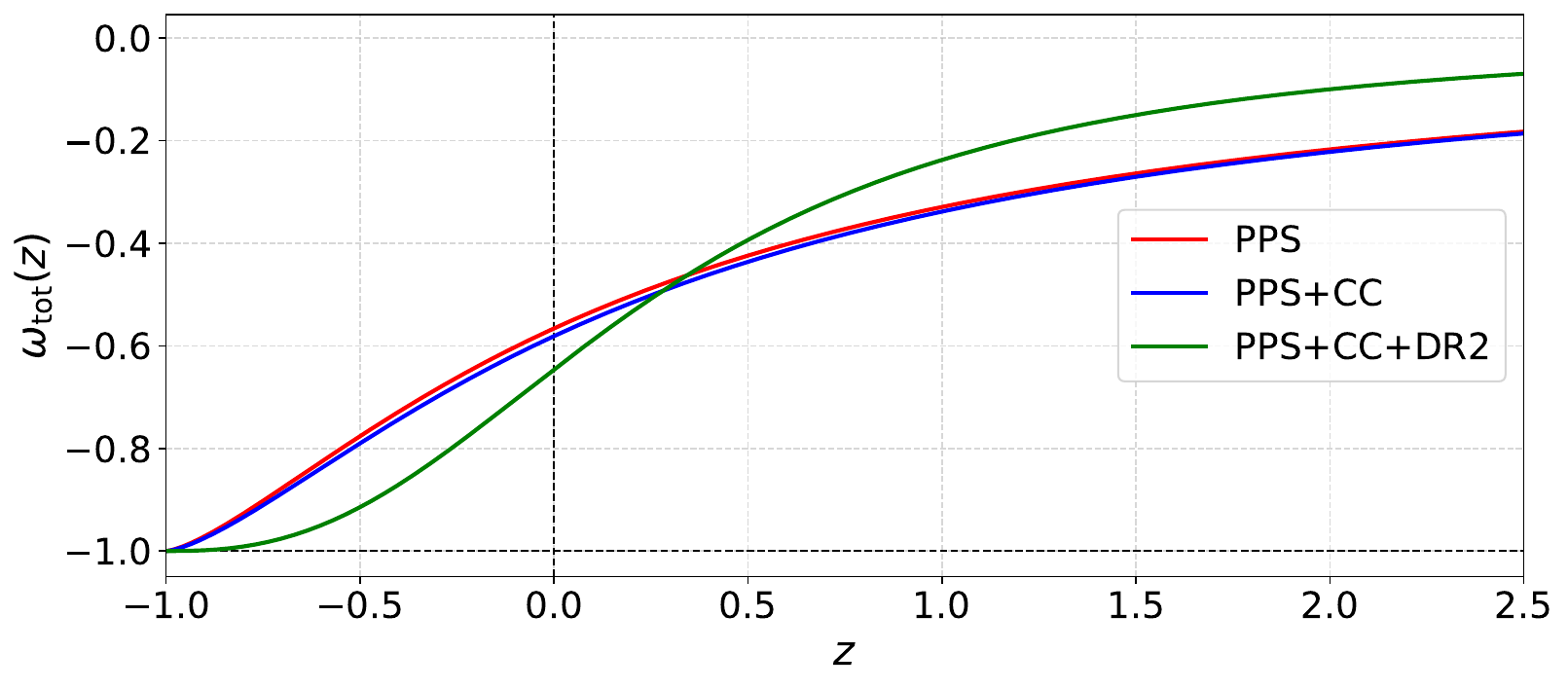}
        \caption{Total equation of state $\omega_{tot}$ vs redshift $z$ for Model 2.}
        \label{fig:3b}
    \end{figure}

Using the above relation together with the model equations, we obtain the evolution of the total EoS parameter for both models. The behavior of $\omega_{\mathrm{tot}}(z)$ is shown in Fig.~\ref{fig:3a} and Fig.~\ref{fig:3b} for Model 1 and Model 2, respectively. It is observed that $\omega_{\mathrm{tot}}(z)$ evolves from values close to zero at higher redshift toward more negative values at lower redshift, indicating a transition from a matter-dominated phase to an accelerated expansion phase.

At the present epoch, the EoS parameter lies within the quintessence region. From Table~IV, the present values are $\omega_0 \approx -0.565$ to $-0.632$ for Model 1 and $\omega_0 \approx -0.566$ to $-0.646$ for Model 2 across different datasets. These values satisfy $\omega_0 < -\frac{1}{3}$, confirming that the Universe is currently undergoing accelerated expansion.
However, in both models, the EoS parameter remains within the quintessence regime and does not cross the phantom boundary ($\omega = -1$), ensuring a physically viable evolution.
\begin{figure}[ht]
        \centering
        \includegraphics[width=1\linewidth]{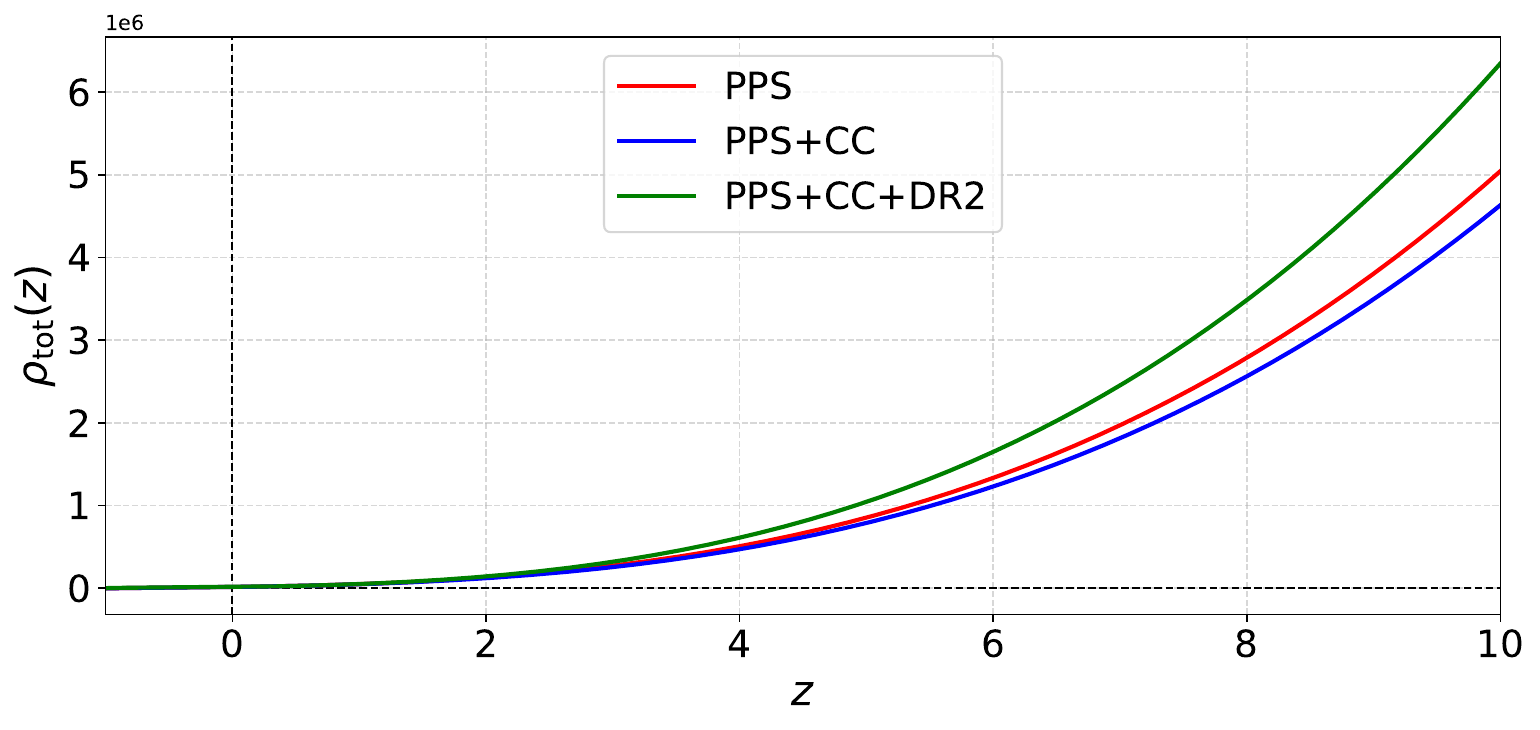}
        \caption{Total energy density $\rho_{tot}$ vs redshift $z$ for Model 1.}
        \label{fig:4a}
    \end{figure}

    \begin{figure}[ht]
        \centering
        \includegraphics[width=1\linewidth]{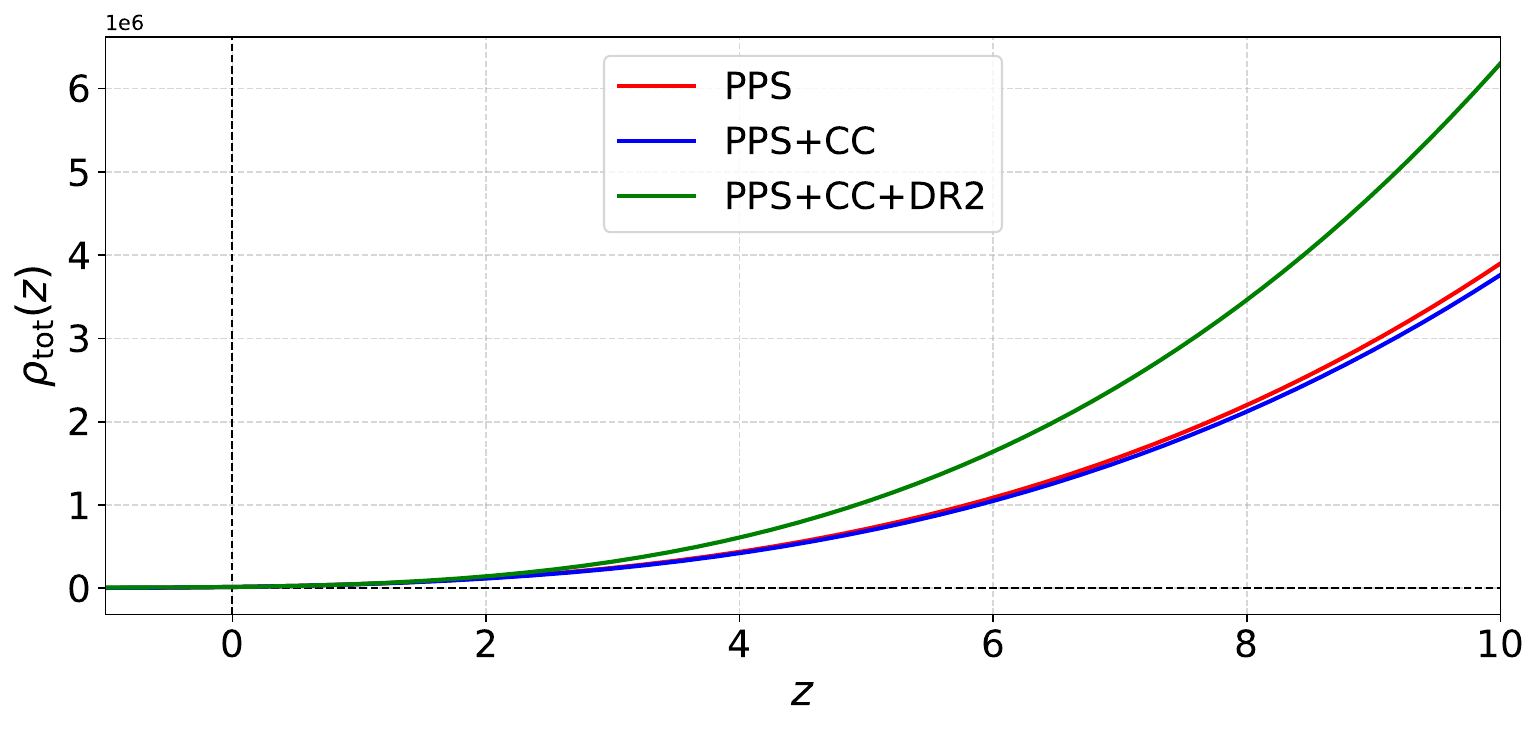}
        \caption{Total energy density $\rho_{tot}$ vs redshift $z$ for Model 2.}
        \label{fig:4b}
    \end{figure}
    
The corresponding total energy density, given by $\rho_{\mathrm{tot}} = 3H^2$, is shown in Fig.~\ref{fig:4a} and Fig.~\ref{fig:4b}. The total energy density increases with redshift due to the growing contribution of matter at earlier times, and decreases smoothly toward lower redshift due to cosmic expansion. This behavior is consistent with the standard cosmological scenario and indicates a stable and well-behaved evolution of the models.

Overall, the evolution of $\omega_{\mathrm{tot}}(z)$ and $\rho_{\mathrm{tot}}(z)$ supports a cosmological picture in which the Universe transitions from a matter-dominated phase in the past to a DE-dominated accelerated phase at present. 
\section{Final Remarks}
\label{sec6}

In this work, we have investigated two phenomenological decaying vacuum models characterized by dynamical evolution of the cosmological term, namely a redshift-dependent form $\Lambda(z) \propto (1+z)^n$ (Model~1) and a Hubble-dependent form $\Lambda \propto H^n$ (Model~2). We have derived observational constraints on the model parameters by performing MCMC analysis using recent datasets, including CC, PPS, and DESI BAO DR2 measurements.

Furthermore, we examined the physical behavior of the models through the deceleration parameter $q(z)$, the transition redshift $z_{\mathrm{tr}}$, and the total equation of state parameter $\omega_{\mathrm{tot}}(z)$. The evolution of the deceleration parameter indicates a smooth transition from an early decelerated phase to the present accelerated expansion of the Universe. The key findings of the present work are summarized as follows:

\begin{itemize}

\item[i)] The joint analysis of PPS + CC + DR2 yields tight constraints on the cosmological parameters. The Hubble constant is found to be $H_0 = 72.93\pm0.23 \mathrm{km\,s^{-1}\,Mpc^{-1}}$ for Model~1 and $H_0 = 73.01\pm 0.26 \mathrm{km\,s^{-1}\,Mpc^{-1}}$ for Model~2, both consistent with local measurements. The matter density parameter $\Omega_{m0}$ decreases systematically with the inclusion of additional datasets, while the evolution parameter is constrained to small positive values ($n \approx 0.30$) for both models.

\item[ii)] The goodness-of-fit analysis shows $\chi^2_r \approx 1.03$ for all dataset combinations, indicating a satisfactory and statistically consistent fit. The small differences in $\chi^2_{\mathrm{min}}$ between Model~1 and Model~2 suggest that neither model is statistically preferred over the other with current observational data.

\item[iii)] Both models consistently predict a negative present value of the deceleration parameter, with $q_0$ lying in the range $\sim -0.35$ to $-0.47$ across different dataset combinations, confirming the accelerated expansion of the Universe. The transition redshift is found to lie in the range $z_{\mathrm{tr}} \sim 0.66$--$1.03$, indicating a transition from deceleration to acceleration at late times.

\item[iv)] The total equation of state parameter remains within the quintessence region ($-1 < \omega_{\mathrm{tot}} < -1/3$) for all dataset combinations and does not cross the phantom boundary, ensuring a physically viable cosmic evolution.

\end{itemize}

\noindent Overall, both decaying vacuum models provide a consistent and viable description of the late-time expansion of the Universe and remain compatible with current observational data. The results show a preference for small positive values of $n$, indicating a mild deviation from the standard $\Lambda$CDM scenario. However, the statistical significance of this deviation is not sufficient to decisively distinguish between a constant vacuum energy and a dynamically evolving vacuum scenario. This highlights the need for future high-precision cosmological observations to further test the nature of dark energy.

\bibliographystyle{utphys}
\bibliography{references}

\end{document}